\documentclass[final,3p,times,twocolumn]{elsarticle}

\usepackage{amssymb}

\usepackage{wallpaper}
\usepackage{dcolumn}
\usepackage{bm}
\usepackage{pdfpages}
\usepackage{graphics}
\usepackage[colorlinks,plainpages=false]{hyperref}
\usepackage{bbding}
\usepackage{amsmath}
\usepackage{upgreek}
\usepackage{multirow}

\journal{Physics Letters B}

\begin{document}
\begin{frontmatter}



\title{Elastic properties of nuclear pasta in a fully three-dimensional geometry}

\author[Adr1,Adr2]{Cheng-Jun~Xia\corref{cor1}}
\ead{cjxia@yzu.edu.cn}
\author[Adr2]{Toshiki Maruyama\corref{cor1}}
\ead{maruyama.toshiki@jaea.go.jp}
\author[Adr3,Adr2]{Nobutoshi Yasutake\corref{cor1}}
\ead{nobutoshi.yasutake@it-chiba.ac.jp}
\author[Adr4]{Toshitaka Tatsumi\corref{cor1}}
\ead{tatsumitoshitaka@gmail.com}
\author[Adr5,Adr6]{Ying-Xun Zhang\corref{cor1}}
\ead{zhyx@ciae.ac.cn}

\cortext[cor1]{Corresponding author.}

\address[Adr1]{Center for Gravitation and Cosmology, College of Physical Science and Technology, Yangzhou University, Yangzhou 225009, China}
\address[Adr2]{Advanced Science Research Center, Japan Atomic Energy Agency, Shirakata 2-4, Tokai, Ibaraki 319-1195, Japan}
\address[Adr3]{Department of Physics, Chiba Institute of Technology (CIT), 2-1-1 Shibazono, Narashino, Chiba, 275-0023, Japan}
\address[Adr4]{Institute of Education, Osaka Sangyo University, 3-1-1 Nakagaito, Daito, Osaka 574-8530, Japan}
\address[Adr5]{China Institute of Atomic Energy, Beijing 102413, People's Republic of China}
\address[Adr6]{Guangxi Key Laboratory Breeding Base of Nuclear Physics and Technology, Guilin 541004, China}

\begin{abstract}
Realistic estimations on the elastic properties of neutron star matter are carried out with a large strain ($\varepsilon \lesssim 0.5$) in the framework of relativistic-mean-field model with Thomas-Fermi approximation, where various crystalline configurations are considered in a fully three-dimensional geometry with reflection symmetry. Our calculation confirms the validity of assuming Coulomb crystals for the droplet phase above neutron drip density, which nonetheless does not work at large densities since the elastic constants are found to be decreasing after reaching their peaks. Similarly, the analytic formulae derived in the incompressible liquid-drop model give excellent description for the rod phase at small densities, which overestimates the elastic constants at larger densities. For slabs, due to the negligence on the variations of their thicknesses, the analytic formulae from liquid-drop model agree qualitatively but not quantitatively with our numerical estimations. By fitting to the numerical results, these analytic formulae are improved by introducing dampening factors. The impacts of nuclear symmetry energy are examined adopting two parameter sets, corresponding to the slope of symmetry energy $L = 41.34$ and 89.39 MeV. Even with the uncertainties caused by the anisotropy in polycrystallines, the elastic properties of neutron star matter obtained with $L = 41.34$ and 89.39 MeV are distinctively different, results in detectable differences in various neutron star activities.
\end{abstract}


\end{frontmatter}

The elastic properties of nuclear pasta play essential roles in understanding the asteroseismology of neutron stars~\cite{Chamel2008_LRR11-10, Caplan2017_RMP89-041002}, which help to constrain nuclear matter properties and unveil the internal compositions of neutron stars, e.g., those in Refs.~\cite{Sotani2012_PRL108-201101, Tsang2012_PRL108-011102}. Significant progresses were made in fixing the elastic properties of outer-crust materials in neutron stars~\cite{Ogata1990_PRA42-4867, Strohmayer1991_ApJ375-679, Horowitz2009_PRL102-191102, Baiko2011_MNRAS416-22, Kozhberov2019_MNRAS486-4473, Chugunov2020_MNRAS500-L17}, which are often extended to the droplet phase above neutron drip density~\cite{Zemlyakov2022}. For inner-crust materials such as the rod and slab phases, their elastic properties were estimated with incompressible liquid-drop model~\cite{Pethick1998_PLB427-7, Pethick2020_PRC101-055802}. Nevertheless, those estimations were obtained neglecting various contributions from free neutrons, strong interaction among nuclei, charge screening, finite sizes of nuclei, neutron superfluidity, and weak reactions, which lead to uncertainties when applying the results to the inner-crust regions of neutron stars. Recently one exception with large-scale classical molecular dynamics simulations contain large amounts of nucleons was carried out for the slab phase~\cite{Caplan2018_PRL121-132701}, which qualitatively confirms the elastic properties of idealized slab phase estimated with incompressible liquid-drop model~\cite{Pethick1998_PLB427-7, Pethick2020_PRC101-055802}. The shape of slabs were modified as well by introducing large deformations, which buckles (with splay deformations) by strain along the direction perpendicular to the slabs or reduces the number of slabs (forming complicated intermediate structures) by strain along them.

In this Letter, we examine the elastic properties of inner-crust materials in various shapes and lattice configurations under more realistic considerations, where the pasta structures are obtained in a three-dimensional geometry with reflection symmetry. The relativistic-mean-field (RMF) model is adopted with both the mean fields and fermion density profiles fixed self-consistently in the framework of Thomas-Fermi approximation~\cite{Okamoto2012_PLB713-284, Okamoto2013_PRC88-025801, Xia2021_PRC103-055812}. In order to extract the elastic properties, the numerical accuracy on the energy per baryon needs to be within $\sim$eV, which are attained by solving the Klein-Gordon equations via fast cosine transformations with fine grid distance ($\lesssim0.35$ fm). The calculation was carried out in Beijing super cloud computing center, where each cluster is equipped with CPUs (AMD 7452@2.35GHz) with 64 cores in total and a flash memory of 256 G. The numerical cost varies with the pasta structures and box sizes, which takes approximately 24 hours for the droplet phase at a fixed average density and configuration. As the oscillation time scale is much larger than weak reactions, the $\beta$-stability condition is always fulfilled in our calculation, while the dripped neutrons, the electron charge screening effect, as well as the complicated nuclear shapes are accounted for. The impacts of symmetry energy are investigated adopting two parameter sets for the isovector channel, which correspond to two different slopes of symmetry energy, i.e., $L=89.39$ (41.34) MeV for Set 0 (1)~\cite{Xia2021_PRC103-055812}.

Based on elasticity theory, the variation of energy density by applying strains $u_{ij}$ on nuclear pasta can be expanded as~\cite{Ogata1990_PRA42-4867}
\begin{equation}
\delta E = \frac{1}{2} \sum_{n=1}^3\sum_{m=1}^3 c_{mn} u_m u_n + 2\sum_{m=4}^6 c_{mm} u_m^2, \label{eq:dE1}
\end{equation}
where the elastic constant $c_{mn}=c_{nm}$ and $u_m = (u_{ij}+u_{ji})/2$. The indices $i$ and $j$ represent the Cartesian components ($x$, $y$, and $z$), while $m$ and $n$ correspond to a transformation ($ij$, $kl$) $\rightarrow$ ($m$, $n$) of the subscripts ($xx$, $yy$, $zz$, $xy$, $yz$, $zx$) $\rightarrow$ (1, 2, 3, 4, 5, 6). For a fixed displacement gradient $u_{ij}$, a droplet at position $\vec{r}$ is moved to a new position $\vec{r}'$ with
\begin{equation}
  r'_i = r_i + \sum_j u_{ij} r_j, \label{eq:drij}
\end{equation}
where in this work we consider a uniform deformation with constant $u_{ij}$. In practice, each droplet is constrained to be centered at $\vec{r}'$ which moves along with the deforming unit cell. For nuclear pasta with cubic symmetry (e.g., BCC and FCC lattices), we have $c_{11}=c_{22}=c_{33}$, $c_{12}=c_{21}=c_{13}=c_{31}=c_{23}=c_{32}$, and $c_{44}=c_{55}=c_{66}$. For rods/tubes that are aligned with $z$-axis, only the terms $c_{11}$ ($=c_{22}$), $c_{12}$, and $c_{44}$ ($=c_{11}/2-c_{12}/2$ for honeycomb configuration) remain nonzero~\cite{Pethick2020_PRC101-055802}, while only $c_{11}$ persists for slabs that are perpendicular to the $x$-axis. The lattice constants $a$, $b$, and $c$ are taken along $x$, $y$, and $z$-axis, respectively.

Carrying out Monte Carlo simulations and assuming point nuclei embedded in a uniform electron background, the elastic constants of BCC and FCC crystals at vanishing temperatures were estimated with~\cite{Ogata1990_PRA42-4867}
\begin{equation}
 \left\{\begin{array}{l}
   \mathrm{BCC:} c_{11}-c_{12}=0.04908 \mu_{0}, \ c_{44}=0.1827 \mu_{0}; \\
   \mathrm{FCC:} c_{11}-c_{12}=0.04132 \mu_{0}, \ c_{44}=0.1852 \mu_{0}. \\
 \end{array}\right. \label{Eq:El_MC}
\end{equation}
Here $\mu_{0}\equiv \alpha n_d Z^2/{R_\mathrm{W}}$ with $n_d$ being the nuclei density, $Z$ the proton number of each nucleus, $R_\mathrm{W} = \left(4\pi n_d/3\right)^{-1/3}$ the Wigner-Seitz (WS) radius, and $\alpha=1/137$ the fine-structure constant. For non-spherical nuclei, the elastic constants of rods in honeycomb configuration and slabs were estimated based on incompressible liquid-drop model~\cite{Pethick1998_PLB427-7, Pethick2020_PRC101-055802}, i.e.,
\begin{eqnarray}
\mathrm{Rods:}&& c_{11}-c_{12}=2c_{44}=2E_\mathrm{C} 10^{2.1\left(u^2-0.3\right)}, \nonumber \\
      && c_{11}+c_{12}=3E_\mathrm{C}, \label{eq:El_ld_Rod} \\
\mathrm{Slabs:}&& c_{11}=6E_\mathrm{C}, \label{eq:El_ld_Slab}
\end{eqnarray}
with $u\equiv{R_d}/{R_\mathrm{W}}$. Note that the surface and Coulomb energy densities are connected by $E_\mathrm{S} = 2E_\mathrm{C}$ for optimized droplet size $R_d$ and WS cell size $R_\mathrm{W}$.

\begin{figure}
\includegraphics[width=\linewidth]{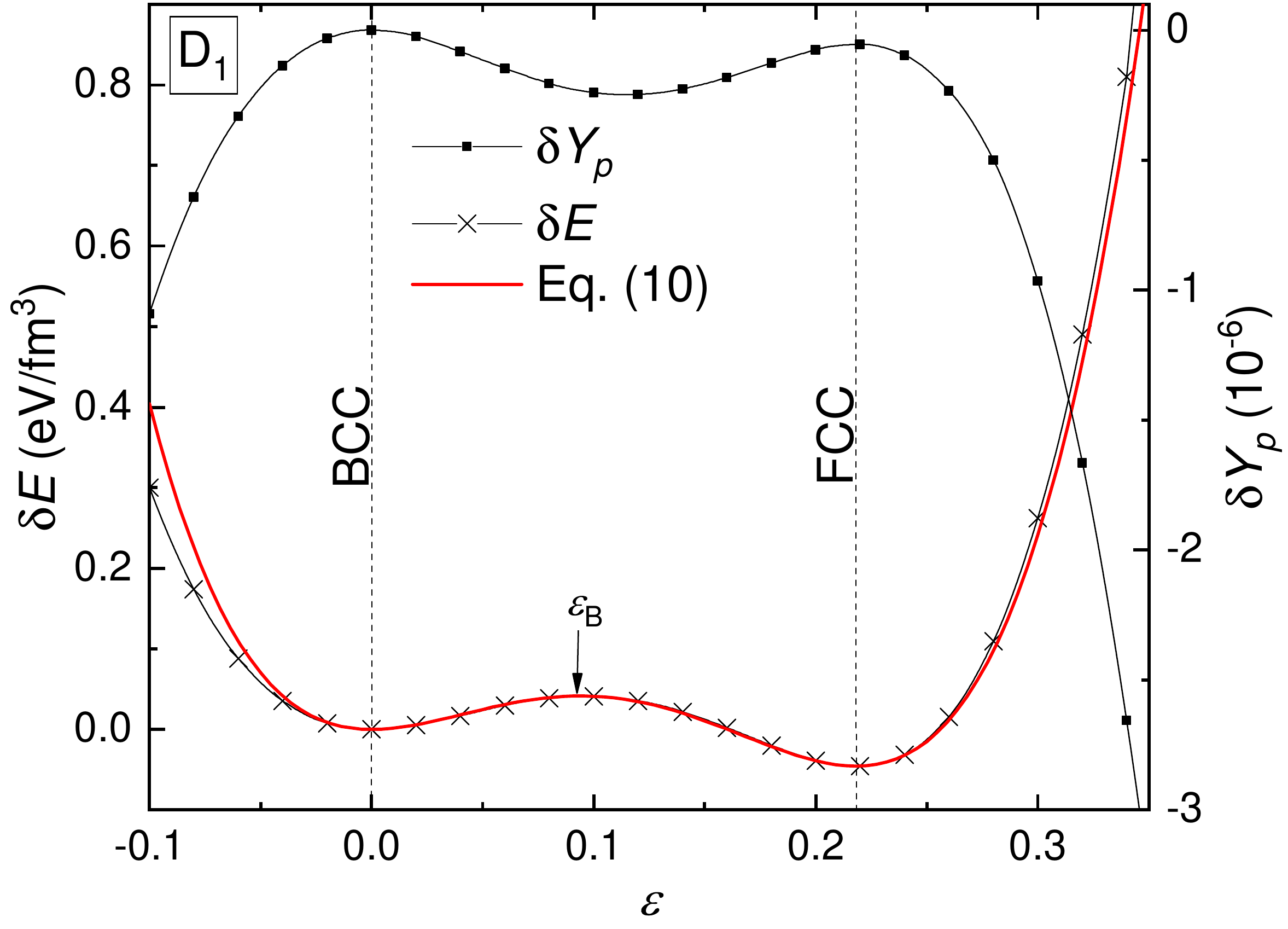}
\caption{\label{Fig:BCC2FCC} Variations of energy density ($\delta E$) and proton fraction ($\delta Y_p$) for the droplet phase in $\beta$-stability as functions of deformation, where parameter Set 0 with $L=89.39$ MeV is adopted and the average baryon number density is fixed at $n_\mathrm{b} = 0.06\ \rm{fm}^{-3}$.}
\end{figure}

\begin{figure}
\includegraphics[width=\linewidth]{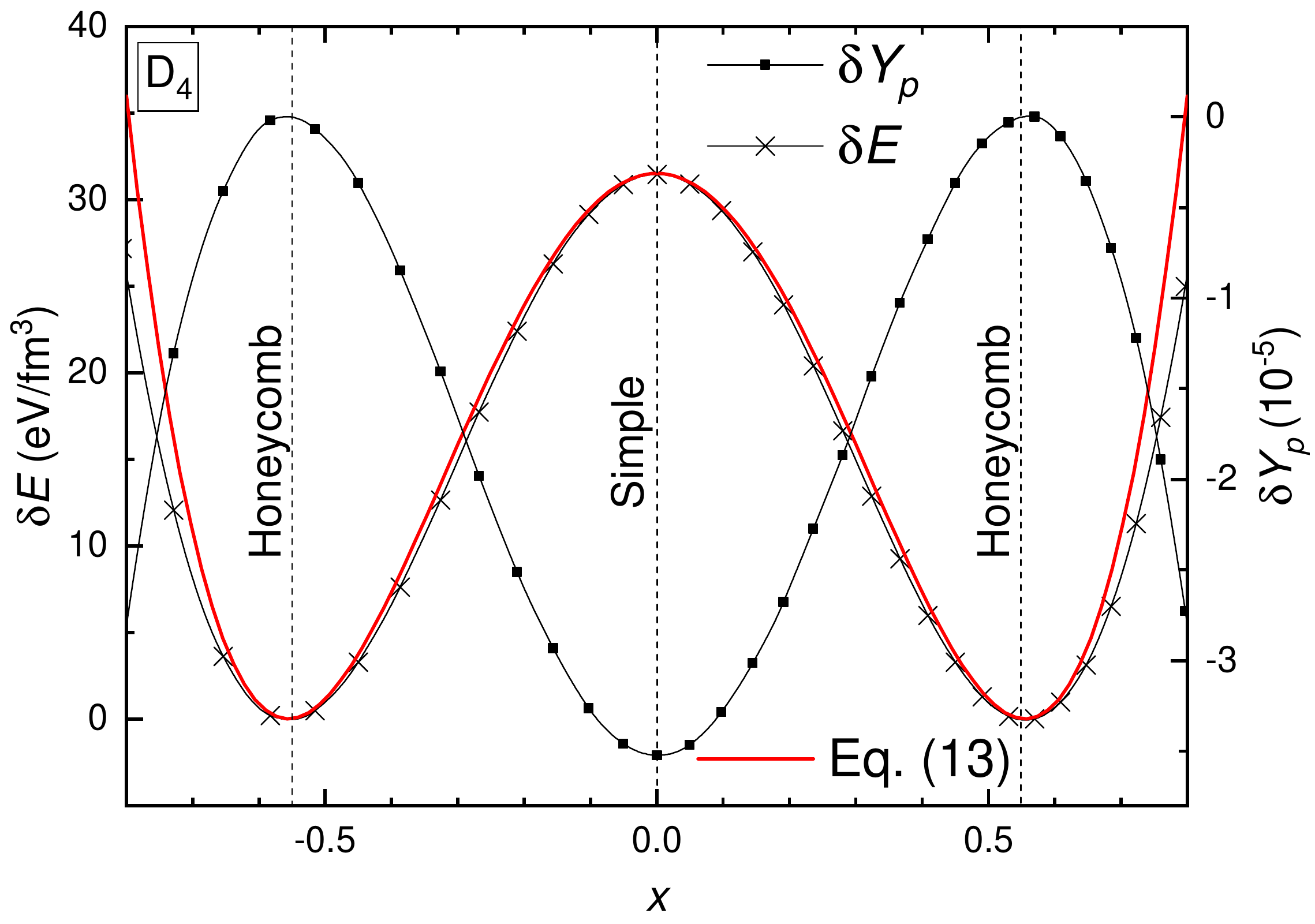}
\caption{\label{Fig:Edformc11mc22} Same as Fig.~\ref{Fig:BCC2FCC} but for the rod phase at $n_\mathrm{b} = 0.06\ \rm{fm}^{-3}$ adopting parameter Set 1 with $L=41.34$ MeV.}
\end{figure}

In this work, to investigate the elastic properties of nuclear pastas in RMF models, we perform volume-preserving deformations similar as in Refs.~\cite{Ogata1990_PRA42-4867, Caplan2018_PRL121-132701}, i.e.,
\begin{eqnarray}
D_1:&& u_{xx} = u_{yy} = -\frac{\varepsilon}{2}, \ \  u_{zz} = \left( 1-\frac{\varepsilon}{2} \right) ^{-2}-1; \label{eq:D1} \\
D_2:&& u_{xy} = u_{yx} = \frac{\varepsilon}{2}, \ \  u_{zz} = \frac{\varepsilon^2}{4-\varepsilon^2};  \label{eq:D2} \\
D_3:&& u_{xx} = -u_{yy} = \frac{\varepsilon}{2}, \ \  u_{zz} = \frac{\varepsilon^2}{4-\varepsilon^2};  \label{eq:D3} \\
D_4:&& u_{xx} = \varepsilon, \ \   u_{yy} = \frac{-\varepsilon}{1+\varepsilon}, \ \  u_{zz} = 0. \label{eq:D4}
\end{eqnarray}
As nuclear droplets are repositioned with Eq.~(\ref{eq:drij}), the structures of unit cells will be altered accordingly while keeping their volumes constant.
The elastic constants $c_{ij}$ can then be estimated by substituting the deformations into Eq.~(\ref{eq:dE1}) and compare the $\varepsilon^2$ terms with respect to the variations of energy density as illustrated in Figs.~\ref{Fig:BCC2FCC} and \ref{Fig:Edformc11mc22}.

For the droplet phase, the BCC lattice (with lattice constants $a=b=c$) can evolve into FCC lattice ($\sqrt{2}a=\sqrt{2}b=c$) by applying the deformation $D_1$ with $\varepsilon=\varepsilon_\mathrm{FCC}=2-2^{5/6}\approx 0.2182$, i.e., along the Bain path~\cite{Bain1924_TAIMME70-25}. The mean fields and energy density are altered by deformations, while the relative contents of protons and neutrons change due to a mismatch between the chemical potentials $\mu_n$ and $\mu_p+\mu_e$. In Fig.~\ref{Fig:BCC2FCC} we present the variations of energy density and proton fraction, where the BCC and FCC lattices represent two local minima separated by the Bain barrier at $\varepsilon=\varepsilon_\mathrm{B} \approx 0.1$, indicating their possible coexistence inside neutron stars. Meanwhile, the variation of proton fraction suggests a chemical potential difference $\delta \mu =\mu_p+\mu_e-\mu_n \approx 1.425 \delta Y_p$ GeV induced by deformation, which enables weak reactions such as $p+e^-\rightarrow n +\nu_e$ and could cause dissipation for oscillating neutron stars. The energy density can be well reproduced by the red-solid curve in Fig.~\ref{Fig:BCC2FCC} within the range $-0.05\lesssim\varepsilon \lesssim 0.34$, which is determined by
\begin{equation}
\delta E = C \left[ \frac{1}{4} \varepsilon^4 - \frac{1}{3} \left( \varepsilon_\mathrm{FCC} + \varepsilon_\mathrm{B} \right) \varepsilon^3+ \frac{1}{2} \varepsilon_\mathrm{FCC} \varepsilon_\mathrm{B} \varepsilon^2 \right] \label{eq:dEBtoF}
\end{equation}
with $C$ being the strength of variation. Note that Eq.~(\ref{eq:dEBtoF}) is obtained with a polynomial expansion to accommodate the two local minima at $\varepsilon = 0$ and $\varepsilon_\mathrm{FCC}$ separated by the local maximum at $\varepsilon = \varepsilon_\mathrm{B}$. The elastic constant $c_{11}-c_{12}$ for the droplet phase is then estimated with
\begin{equation}
 c_{11}-c_{12} =
 \left\{\begin{array}{l}
   2C\varepsilon_\mathrm{FCC} \varepsilon_\mathrm{B}/3, \\
   C \varepsilon_\mathrm{FCC} \left( \varepsilon_\mathrm{FCC} -2 \right)^2 \left(\varepsilon_\mathrm{FCC}-\varepsilon_\mathrm{B}\right)/6, \\
 \end{array}\right.
 \begin{array}{l}\text{BCC}\\\text{FCC}\\ \end{array}.  \label{Eq:c11m2_drop}
\end{equation}
The coefficients $C$ and $\varepsilon_\mathrm{B}$, in turn, can be fixed for given $c_{11}-c_{12}$, which is useful for larger deformations with Eq.~(\ref{eq:dEBtoF}). The elastic constant $c_{44}$ is fixed by applying the deformation $D_2$ and examine the energy variation. Equivalently, we rotate the nuclear pasta $45^\circ$ along $z$-axis and apply the deformation $D_3$, where the variations of energy density at $\varepsilon\lesssim 0.18$ are well reproduced with
\begin{equation}
  \delta E=c_{44}\varepsilon^2/2. \label{Eq:c44_drop}
\end{equation}

\begin{figure}
\includegraphics[width=\linewidth]{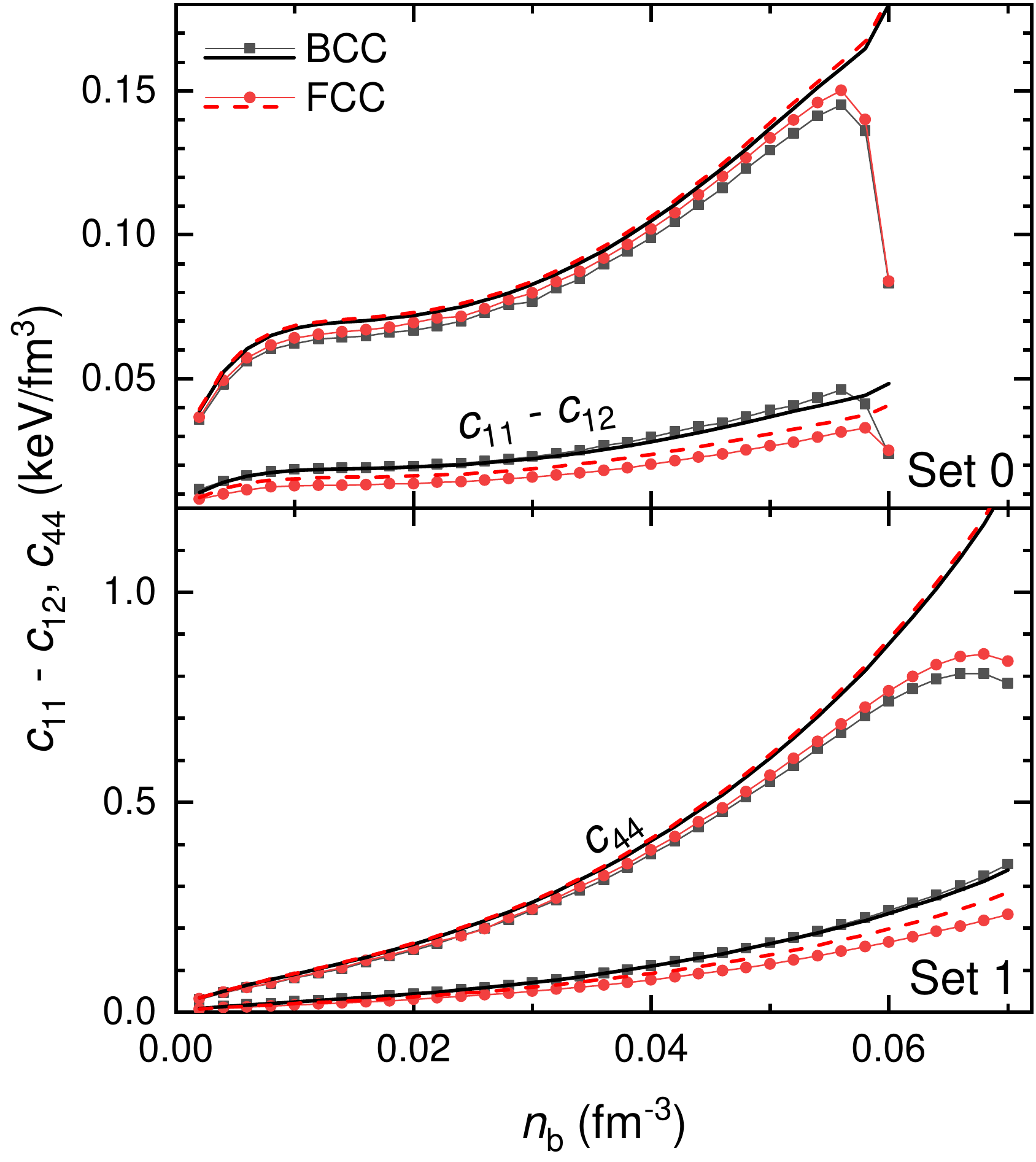}
\caption{\label{Fig:ElasDrop} Elastic properties of droplet phases in BCC and FCC configurations in neutron stars, where the lines represent the values fixed by Eq.~(\ref{Eq:El_MC}) and symbols by Eqs.~(\ref{Eq:c11m2_drop}-\ref{Eq:c44_drop}).}
\end{figure}

Figure~\ref{Fig:ElasDrop} shows the obtained elastic constants of the droplet phase, which generally increase with density. Our numerical estimations with Eqs.~(\ref{Eq:c11m2_drop}-\ref{Eq:c44_drop}) confirms the validity of assuming Coulomb crystals for the droplet phase above neutron drip density, where Eq.~(\ref{Eq:El_MC}) predicts similar values at small densities. Nevertheless, there exist slight discrepancies, which are attributed to various contributions such as the dripped neutrons, charge screening effects, and small deviations from the spherical shapes of droplets~\cite{Zemlyakov2022}. Note that the fulfillment of $\beta$-stability condition barely affects the elastic properties in spite of the variations in the proton fraction and Coulomb energy under deformation. As density increases, the elastic constants start to decrease after reaching their peaks, which is not predicted by Eq.~(\ref{Eq:El_MC}). The softening of the droplet phase is mainly attributed to the increment of neutron gas density, where the liquid-gas interface becomes less evident. These effects can be partially accounted for by subtracting the background proton number density $n_p(R_\mathrm{W})$ in Eq.~(\ref{Eq:El_MC}) by replacing $Z$ with $Z_\mathrm{droplet}=Z-4\pi{R_\mathrm{W}}^3 n_p(R_\mathrm{W})/3$, which predicts similar trends with slight deviations for Set 0. Note that despite its dominance at smaller densities, the Coulomb interaction gives negative contributions to the elastic constants at large densities~\cite{Xia2022_PRD106-063020}. For the droplet phase obtained with Set 1, replacing $Z$ with $Z_\mathrm{droplet}$ does not improve Eq.~(\ref{Eq:El_MC}) since $n_p(R_\mathrm{W})=0$, which predicts distinctively different trends for $c_{44}$ as indicated in the lower panel of Fig.~\ref{Fig:ElasDrop}. In this case, as the distance between droplets decreases with density $n_\mathrm{b}$, quadrupole electrostatic potential will induce deformations and reduce $c_{44}$~\cite{Zemlyakov2022}. If we fit our numerical results, Eq.~(\ref{Eq:El_MC}) should be modified by introducing dampening factors as indicated in Table.~\ref{tab:damp}. The slope of symmetry energy $L$ plays an essential role on the elastic properties of nuclear pasta. On the one hand, the phase diagram is affected by $L$ with only droplet phase persists up to the core-crust transition density $n_\mathrm{t}=0.061$ fm${}^{-3}$ for Set 0 ($L = 89.39$ MeV), while deformed nuclei are formed at $0.069\lesssim n_\mathrm{b} \lesssim 0.089$ fm${}^{-3}$ for Set 1 ($L = 41.34$ MeV). On the other hand, with modifications to the microscopic structures of nuclear pasta, the elastic constants of the droplet phase with $L = 41.34$ MeV are almost ten times larger than those with $L = 89.39$ MeV. Note that the symmetry energy becomes larger at subsaturation densities for smaller $L$, which enhances the proton-neutron interactions at $L = 41.34$ MeV and leads to larger proton numbers $Z$ of nuclei.

For the rod/tube phases, the simple configuration ($b=a$) can evolve into the honeycomb one ($b=\sqrt{3}a$) by applying the deformation $D_4$ with $\varepsilon =3^{\pm1/4}-1$, where in Fig.~\ref{Fig:Edformc11mc22} we present the variation of energy density and proton fraction with respect to deformation with $x\equiv{\varepsilon \left( \varepsilon+2 \right) }/{(\varepsilon+1)} = (a-b)/\sqrt{ab}$. It is evident that $\delta E(x)$ is symmetric under reflection at $x=0$, while the simple configuration corresponds to a local maximum and is thus unstable against decaying into the honeycomb one under deformation $D_4$. Similar to Fig.~\ref{Fig:BCC2FCC}, the nonzero values of $\delta Y_p$ are caused by a chemical potential difference $\delta \mu \approx 911 \delta Y_p$ MeV under deformation. The variation of energy density with $|x|\lesssim 0.65$ ($-0.27\lesssim\varepsilon \lesssim 0.37$) can be well described by the following polynomial
\begin{equation}
\delta E = \frac{3}{8} c_{44} \left[x^2-\frac{4}{\sqrt{3}}+2 \right]^2, \label{eq:dESimptoHoney}
\end{equation}
which reproduces the local minima and maximum in correspondence to the honeycomb and simple configurations. The elastic constants of the honeycomb configuration can then be extract with $c_{11} - c_{12} = 2 c_{44}$. The last independent elastic constant $c_{11}$ ($=c_{22}$) can be estimated by applying the deformation $D_4$ exchanging $y$ and $z$ axis with $u_{yy}=0$, where the variation of energy density is determined by
\begin{equation}
  \delta E=c_{11}\varepsilon^2/2. \label{Eq:c11_rod}
\end{equation}
For the slab phase, the elastic constant $c_{11}$ is obtained in the same manner.

\begin{figure}
\includegraphics[width=\linewidth]{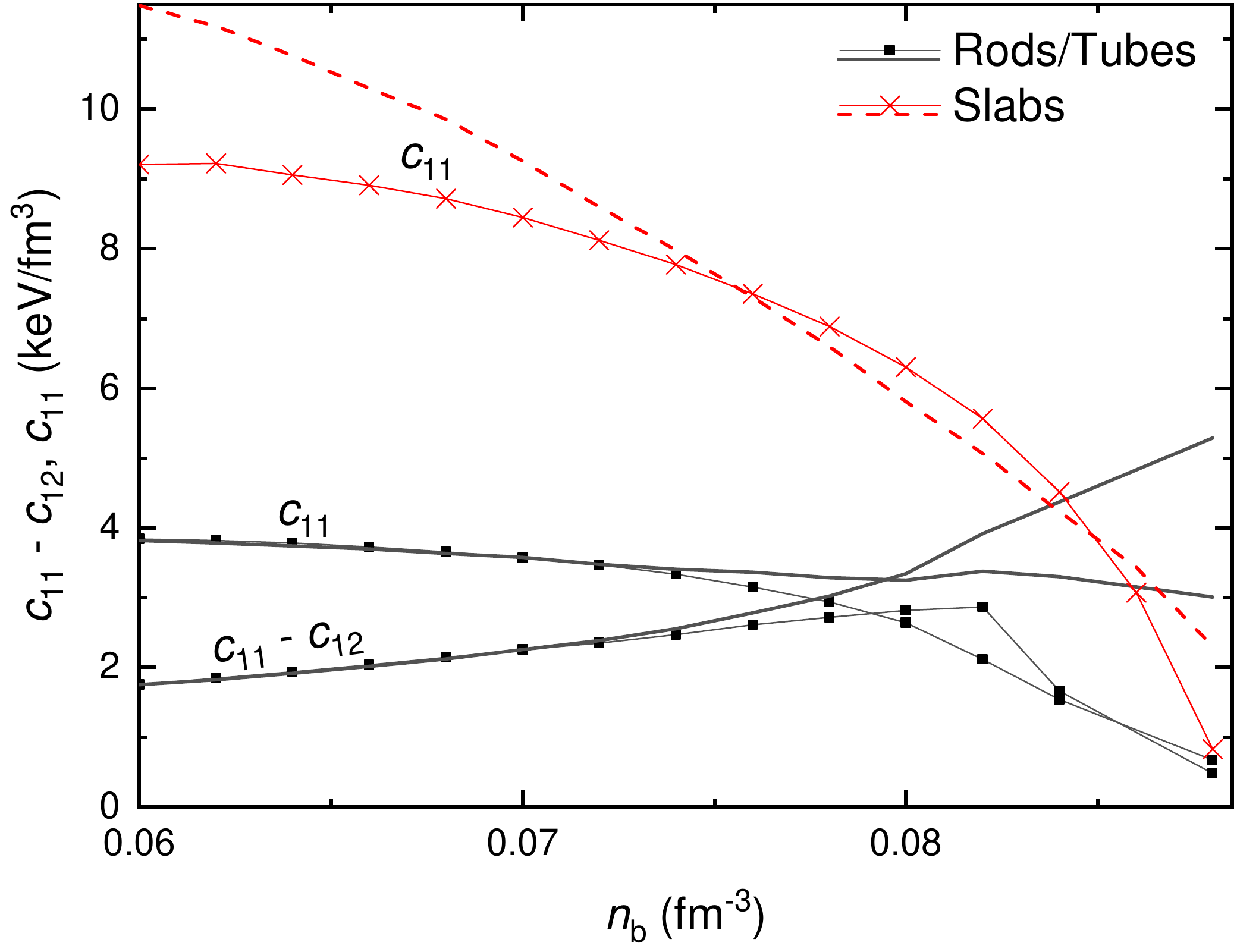}
\caption{\label{Fig:ElasPasta} Elastic properties for the rod/tube and slab phases in neutron stars, where the lines represent the values fixed by Eqs.~(\ref{eq:El_ld_Rod}-\ref{eq:El_ld_Slab}) and symbols by Eqs.~(\ref{eq:dESimptoHoney}-\ref{Eq:c11_rod}). The parameter Set 1 with $L=41.34$ MeV is adopted.}
\end{figure}

\begin{table}
\caption{\label{tab:damp} Analytical formulae that reproduce our numerical results by introducing dampening factors to Eqs.~(\ref{Eq:El_MC}), (\ref{eq:El_ld_Rod}) and (\ref{eq:El_ld_Slab}). Here $\mu_{0}'\equiv \alpha n_d Z_\mathrm{droplet}^2/{R_\mathrm{W}}$ and $u\equiv{R_d}/{R_\mathrm{W}}$.}
\begin{tabular}{c|c|c|c} \hline \hline
                  &       &  $c_{11}-c_{12}$      &    $c_{44}$  \\  \hline
\multirow{2}{*}{Droplets} & BCC   &  $0.04908 \mu_{0}'$     &   $0.1705 \mu_{0}'/10^{10u^8}$     \\
                  & FCC   &  $0.03512 \mu_{0}'$     &   $0.1740 \mu_{0}'/10^{9u^8}$ \\ \hline
  Rods &  \multicolumn{2}{c|}{$c_{11}+c_{12}=3E_\mathrm{C}/10^{3u^8}$} &   $E_\mathrm{C} 10^{2.1\left(u^2-0.3\right)-3u^8}$     \\  \hline
\multicolumn{2}{c|}{Slabs}           &   \multicolumn{2}{c}{$c_{11}=6E_\mathrm{C} 10^{0.55u-10u^8-0.19}$}  \\ \hline
\end{tabular}
\end{table}

Figure~\ref{Fig:ElasPasta} gives the elastic properties of slabs and rods/tubes in honeycomb configuration, which are much stronger than the droplet phase indicated in Fig.~\ref{Fig:ElasDrop}. Nevertheless, the other elastic constants vanish for the rod/tube and slab phases, while those of the droplet phase persist and fulfill cubic symmetry. For the elastic properties of rods in honeycomb configuration, the incompressible liquid-drop model gives excellent description at small densities, while at larger densities Eq.~(\ref{eq:El_ld_Rod}) overestimates the elastic constants in comparison with our numerical estimations indicated by the symbols connected by solid curves. For the slab phase, there are slight differences between the results obtained with Eq.~(\ref{eq:El_ld_Slab}) and our numerical estimations with Eq.~(\ref{Eq:c11_rod}). It is important to note that the thickness of a slab was assumed constant in deriving Eq.~(\ref{eq:El_ld_Slab}) in the incompressible liquid-drop model~\cite{Pethick1998_PLB427-7}, while according to our calculation the density as well as the droplet size $R_d$ increases with $\varepsilon$ under deformation $D_4$ with $u_{yy}=0$. In general, the elastic constants in Fig.~\ref{Fig:ElasPasta} decrease with density, which are expected to vanish at $n_\mathrm{b}\approx n_\mathrm{t}$ as the neutron star matter becomes uniform. By fitting to our numerical results in Fig.~\ref{Fig:ElasPasta}, Eqs.~(\ref{eq:El_ld_Rod}) and (\ref{eq:El_ld_Slab}) can be improved by introducing dampening factors as indicated in Table.~\ref{tab:damp}. We note that the elastic constants are generally proportional to the Coulomb energy density, which are expected to be increasing with respect to the surface tension of nuclear matter. Meanwhile, according to the numerical estimations in Ref.~\cite{Ogata1990_PRA42-4867}, the elastic constants are expected to decrease with temperature $T$ and start to vanish as $T$ approaches to the melting temperature. To estimate the elastic properties of nuclear pasta at large temperatures, a detailed calculation should be carried out in our future works.

\begin{figure}
\includegraphics[width=\linewidth]{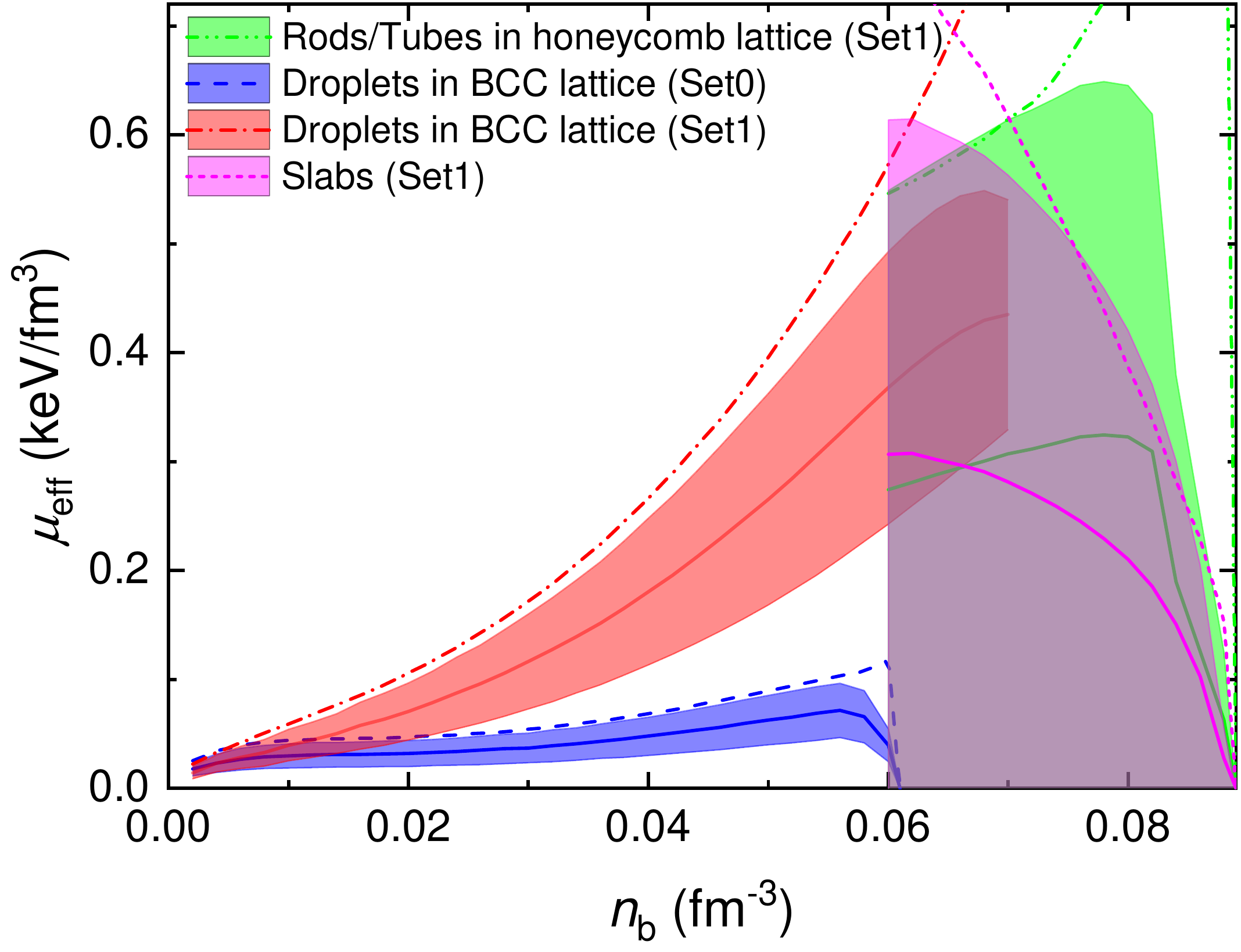}
\caption{\label{Fig:mueff} Effective shear modules of polycrystalline nuclear pastas, where the shaded regions indicate the possible values enclosed by those predicted by the Voigt (upper limit) and Reuss (lower limit) theories. The solid curves located at the center are fixed by arithmetical Hill method~\cite{Hill1952_PPSAA65-349}, while the dashed curves are obtained with Voigt theory using Eqs.~(\ref{Eq:El_MC}-\ref{eq:El_ld_Slab}) for elastic constants.}
\end{figure}

Finally, we should mention that the elastic constants presented here are for single crystals, while in practice one expects the nuclear pastas to form polycrystalline structures inside neutron stars. Based on the extent of anisotropy in polycrystallines, the elastic properties vary within the bounds predicted by the Voigt and Reuss theories~\cite{Hill1952_PPSAA65-349}. By averaging over all possible lattice orientations for the stress in a given strain, the effective shear module in the Voigt theory is given by
\begin{equation}
15 \mu_\mathrm{eff}^\mathrm{V} = c_{11}+c_{22}+c_{33}-c_{12}-c_{13}-c_{23} + 3 (c_{44}+c_{55}+c_{66}).  \label{eq:Voigt_mu}
\end{equation}
Meanwhile, the effective shear module in the Reuss theory is fixed by averaging the strain at a given stress, which gives
\begin{equation}
 \left\{\begin{array}{l}
   \mathrm{droplets:}\  5/\mu_\mathrm{eff}^\mathrm{R} = 4/(c_{11}-c_{12}) + 3/c_{44},\\
   \mathrm{rods/tubes/slabs:}\  \mu_\mathrm{eff}^\mathrm{R} = 0. \\
 \end{array}\right. \label{Eq:Reuss_mu}
\end{equation}

The obtained results are presented in Fig.~\ref{Fig:mueff}, where $\mu_\mathrm{eff}^\mathrm{V}$ and $\mu_\mathrm{eff}^\mathrm{R}$ are indicated by the upper and lower bounds of the shaded regions. Their average values $\mu_\mathrm{eff} = (\mu_\mathrm{eff}^\mathrm{V} + \mu_\mathrm{eff}^\mathrm{R})/2$ according to the arithmetical Hill method~\cite{Hill1952_PPSAA65-349} are indicated by the solid curves. Note that some of the nuclear pasta structures are unstable according to our previous study~\cite{Xia2021_PRC103-055812}, which are included as well considering their possible coexistence~\cite{Xia2022_PRD106-063020} with the actual effective shear module being a combination of the values indicated in Fig.~\ref{Fig:mueff}. Similar to the elastic constants in Figs.~\ref{Fig:ElasDrop} and \ref{Fig:ElasPasta}, adopting the formulae derived in Monte Carlo simulations and the incompressible liquid-drop model generally overestimate the effective shear modules in the Voigt theory, while we find $\mu_\mathrm{eff}$ slowly decreases after reaching its peak and vanishes at the core-crust transition densities $n_\mathrm{t}$. In particular, our estimation shows that $\mu_\mathrm{eff}$ decreases at large enough densities even before the non-spherical nuclei emerge. For perfect rod/tube/slab phases, $\mu_\mathrm{eff}^\mathrm{R}$ vanishes due to the vanishing elastic constants, which may not be the case if they take nonzero values with modulations~\cite{Caplan2018_PRL121-132701, Pethick2020_PRC101-055802}. Even with the uncertainties due to the anisotropy in polycrystallines, the effective shear modules predicted by Set 1 with $L = 41.34$ MeV are almost ten times larger than those with $L = 89.39$ MeV, causing detectable differences in various neutron star activities, e.g., the frequencies of quasiperiodic oscillations in magnetars~\cite{Sotani2012_PRL108-201101} and the short gamma ray burst precursors due to the resonant excitation of neutron star modes in binary systems~\cite{Tsang2012_PRL108-011102}.

\section*{ACKNOWLEDGMENTS}
C.-J. X. would like to thank Dr. Yong Gao, Prof. Yi-Qiu Ma, Prof. Takashi Nakatsukasa, Prof. Xiao-Ping Zheng, and Prof. En-Ping Zhou for fruitful discussions. This work was supported by the National SKA Program of China (Grant No.~2020SKA0120300), National Natural Science Foundation of China (Grants No.~12275234 and No.~11875323), and JSPS KAKENHI (Grants No.~20K03951 and No.~20H04742).




\begin{thebibliography}{20}
\expandafter\ifx\csname natexlab\endcsname\relax\def\natexlab#1{#1}\fi
\providecommand{\bibinfo}[2]{#2}
\ifx\xfnm\relax \def\xfnm[#1]{\unskip,\space#1}\fi
\bibitem[{Chamel and Haensel(2008)}]{Chamel2008_LRR11-10}
\bibinfo{author}{N.~Chamel}, \bibinfo{author}{P.~Haensel},
  \bibinfo{journal}{Living Rev. Rel.} \bibinfo{volume}{11}
  (\bibinfo{year}{2008}) \bibinfo{pages}{10}.
\bibitem[{Caplan and Horowitz(2017)}]{Caplan2017_RMP89-041002}
\bibinfo{author}{M.~E. Caplan}, \bibinfo{author}{C.~J. Horowitz},
  \bibinfo{journal}{Rev. Mod. Phys.} \bibinfo{volume}{89}
  (\bibinfo{year}{2017}) \bibinfo{pages}{041002}.
\bibitem[{Sotani et~al.(2012)Sotani, Nakazato, Iida, and
  Oyamatsu}]{Sotani2012_PRL108-201101}
\bibinfo{author}{H.~Sotani}, \bibinfo{author}{K.~Nakazato},
  \bibinfo{author}{K.~Iida}, \bibinfo{author}{K.~Oyamatsu},
  \bibinfo{journal}{Phys. Rev. Lett.} \bibinfo{volume}{108}
  (\bibinfo{year}{2012}) \bibinfo{pages}{201101}.
\bibitem[{Tsang et~al.(2012)Tsang, Read, Hinderer, Piro, and
  Bondarescu}]{Tsang2012_PRL108-011102}
\bibinfo{author}{D.~Tsang}, \bibinfo{author}{J.~S. Read},
  \bibinfo{author}{T.~Hinderer}, \bibinfo{author}{A.~L. Piro},
  \bibinfo{author}{R.~Bondarescu}, \bibinfo{journal}{Phys. Rev. Lett.}
  \bibinfo{volume}{108} (\bibinfo{year}{2012}) \bibinfo{pages}{011102}.
\bibitem[{Ogata and Ichimaru(1990)}]{Ogata1990_PRA42-4867}
\bibinfo{author}{S.~Ogata}, \bibinfo{author}{S.~Ichimaru},
  \bibinfo{journal}{Phys. Rev. A} \bibinfo{volume}{42} (\bibinfo{year}{1990})
  \bibinfo{pages}{4867--4870}.
\bibitem[{{Strohmayer} et~al.(1991){Strohmayer}, {Ogata}, {Iyetomi},
  {Ichimaru}, and {van Horn}}]{Strohmayer1991_ApJ375-679}
\bibinfo{author}{T.~{Strohmayer}}, \bibinfo{author}{S.~{Ogata}},
  \bibinfo{author}{H.~{Iyetomi}}, \bibinfo{author}{S.~{Ichimaru}},
  \bibinfo{author}{H.~M. {van Horn}}, \bibinfo{journal}{Astrophys. J.}
  \bibinfo{volume}{375} (\bibinfo{year}{1991}) \bibinfo{pages}{679}.
\bibitem[{Horowitz and Kadau(2009)}]{Horowitz2009_PRL102-191102}
\bibinfo{author}{C.~J. Horowitz}, \bibinfo{author}{K.~Kadau},
  \bibinfo{journal}{Phys. Rev. Lett.} \bibinfo{volume}{102}
  (\bibinfo{year}{2009}) \bibinfo{pages}{191102}.
\bibitem[{Baiko(2011)}]{Baiko2011_MNRAS416-22}
\bibinfo{author}{D.~A. Baiko}, \bibinfo{journal}{Mon. Not. R. Astron. Soc.}
  \bibinfo{volume}{416} (\bibinfo{year}{2011}) \bibinfo{pages}{22--31}.
\bibitem[{Kozhberov(2019)}]{Kozhberov2019_MNRAS486-4473}
\bibinfo{author}{A.~A. Kozhberov}, \bibinfo{journal}{Mon. Not. R. Astron. Soc.}
  \bibinfo{volume}{486} (\bibinfo{year}{2019}) \bibinfo{pages}{4473--4478}.
\bibitem[{Chugunov(2020)}]{Chugunov2020_MNRAS500-L17}
\bibinfo{author}{A.~I. Chugunov}, \bibinfo{journal}{Mon. Not. R. Astron. Soc.}
  \bibinfo{volume}{500} (\bibinfo{year}{2020}) \bibinfo{pages}{L17--L21}.
\bibitem[{Zemlyakov and Chugunov(2022)}]{Zemlyakov2022}
\bibinfo{author}{N.~A. Zemlyakov}, \bibinfo{author}{A.~I. Chugunov}, \bibinfo{journal}{Mon. Not. R. Astron. Soc.}
  \bibinfo{volume}{518} (\bibinfo{year}{2022}) \bibinfo{pages}{3813--3819}.
\bibitem[{Pethick and Potekhin(1998)}]{Pethick1998_PLB427-7}
\bibinfo{author}{C.~Pethick}, \bibinfo{author}{A.~Potekhin},
  \bibinfo{journal}{Phys. Lett. B} \bibinfo{volume}{427} (\bibinfo{year}{1998})
  \bibinfo{pages}{7 -- 12}.
\bibitem[{Pethick et~al.(2020)Pethick, Zhang, and
  Kobyakov}]{Pethick2020_PRC101-055802}
\bibinfo{author}{C.~J. Pethick}, \bibinfo{author}{Z.-W. Zhang},
  \bibinfo{author}{D.~N. Kobyakov}, \bibinfo{journal}{Phys. Rev. C}
  \bibinfo{volume}{101} (\bibinfo{year}{2020}) \bibinfo{pages}{055802}.
\bibitem[{Caplan et~al.(2018)Caplan, Schneider, and
  Horowitz}]{Caplan2018_PRL121-132701}
\bibinfo{author}{M.~E. Caplan}, \bibinfo{author}{A.~S. Schneider},
  \bibinfo{author}{C.~J. Horowitz}, \bibinfo{journal}{Phys. Rev. Lett.}
  \bibinfo{volume}{121} (\bibinfo{year}{2018}) \bibinfo{pages}{132701}.
\bibitem[{Okamoto et~al.(2012)Okamoto, Maruyama, Yabana, and
  Tatsumi}]{Okamoto2012_PLB713-284}
\bibinfo{author}{M.~Okamoto}, \bibinfo{author}{T.~Maruyama},
  \bibinfo{author}{K.~Yabana}, \bibinfo{author}{T.~Tatsumi},
  \bibinfo{journal}{Phys. Lett. B} \bibinfo{volume}{713} (\bibinfo{year}{2012})
  \bibinfo{pages}{284--288}.
\bibitem[{Okamoto et~al.(2013)Okamoto, Maruyama, Yabana, and
  Tatsumi}]{Okamoto2013_PRC88-025801}
\bibinfo{author}{M.~Okamoto}, \bibinfo{author}{T.~Maruyama},
  \bibinfo{author}{K.~Yabana}, \bibinfo{author}{T.~Tatsumi},
  \bibinfo{journal}{Phys. Rev. C} \bibinfo{volume}{88} (\bibinfo{year}{2013})
  \bibinfo{pages}{025801}.
\bibitem[{Xia et~al.(2021)Xia, Maruyama, Yasutake, Tatsumi, and
  Zhang}]{Xia2021_PRC103-055812}
\bibinfo{author}{C.-J. Xia}, \bibinfo{author}{T.~Maruyama},
  \bibinfo{author}{N.~Yasutake}, \bibinfo{author}{T.~Tatsumi},
  \bibinfo{author}{Y.-X. Zhang}, \bibinfo{journal}{Phys. Rev. C}
  \bibinfo{volume}{103} (\bibinfo{year}{2021}) \bibinfo{pages}{055812}.
\bibitem[{Bain(1924)}]{Bain1924_TAIMME70-25}
\bibinfo{author}{E.~C. Bain}, \bibinfo{journal}{Trans. Am. Inst. Min. metall.
  Engrs.} \bibinfo{volume}{70} (\bibinfo{year}{1924}) \bibinfo{pages}{25}.
\bibitem[{Xia et~al.(2022)Xia, Maruyama, Yasutake, and
  Tatsumi}]{Xia2022_PRD106-063020}
\bibinfo{author}{C.-J. Xia}, \bibinfo{author}{T.~Maruyama},
  \bibinfo{author}{N.~Yasutake}, \bibinfo{author}{T.~Tatsumi},
  \bibinfo{journal}{Phys. Rev. D} \bibinfo{volume}{106} (\bibinfo{year}{2022})
  \bibinfo{pages}{063020}.
\bibitem[{Hill(1952)}]{Hill1952_PPSAA65-349}
\bibinfo{author}{R.~Hill}, \bibinfo{journal}{Proc. Phys. Soc. A}
  \bibinfo{volume}{65} (\bibinfo{year}{1952}) \bibinfo{pages}{349--354}.

\end{thebibliography}

\end{document}